\documentclass[twocolumn,showpacs,preprintnumbers,amsmath,date,amssymb,prl]{revtex4}

\usepackage{graphicx}
\usepackage{dcolumn}
\usepackage{bm}
\usepackage{natbib}

\newcommand{\be}[1]{\begin{equation}\label{eq:#1}}
\newcommand{\ee}{\end{equation}}
\newcommand{\bea}{\begin{eqnarray}}
\newcommand{\eea}{\end{eqnarray}}
\newcommand{\bt}{\textbf}

\newcommand{\phd}{\phantom{\dag}}
\newcommand{\ph}{\phantom{.}}
\newcommand{\up}{^{\phd}}
\newcommand{\noi}{\noindent}
\newcommand{\no}{\nonumber}

\begin{document}
\def\v#1{{\bf #1}}


\title{Chirality induced tilted-hill giant Nernst signal}

\author{P. Kotetes}\email{pkotetes@central.ntua.gr}
\author{G. Varelogiannis}
\affiliation{Department of Physics, National Technical University
of Athens, GR-15780 Athens, Greece}

\vskip 1cm

\begin{abstract}
We reveal a novel source of giant Nernst response exhibiting strong non-linear temperature and magnetic field dependence including the mysterious tilted-hill temperature profile observed in a pleiad of materials. The phenomenon results directly from the formation of a chiral ground state, e.g. a chiral d-density wave, which is compatible with the eventual observation of diamagnetism and is distinctly different from the usual quasiparticle and vortex Nernst mechanisms. Our picture provides a unified understanding of the anomalous thermoelectricity observed in materials as diverse as the hole-doped cuprates and heavy-fermion compounds like ${\rm URu_2Si_2}$. 
\end{abstract}

 \maketitle


The quintessence of condensed matter physics resides on finding an elegant way to describe the emergent universality in diverse quantum states of matter, ranging for instance from the cuprate and heavy fermion compounds even up to the novel high-Tc iron-based superconductors. The stri\-king common characteristic of all these mate\-rials is a giant Nernst signal \cite{cuprates,NernstFeAs,CeCoIn5,URu2Si2}, typically of the order of $\mu V/K$, exhibiting strong non-linear dependence on temperature and occasionally on the applied field. In the majority of these compounds, the origin of the anomalous thermoelectricity still remains a puzzle, mainly due to the difficulty of explaining the large Nernst magnitude, sign and temperature evolution, on an equal footing.

Up to now, the enhanced Nernst signal has been attributed to a quasi-particle \cite{quasiparticles} or a vortex origin \cite{VortexScenarios}. In a single band quasi-particle picture it is quite difficult to obtain a large Nernst signal due to Sondheimer cancelation, that however relaxes in a multi-band system \cite{NernstMulti}. In this case one may obtain an enhanced thermoelectric response that, nevertheless, depends strongly on the detailed band structure characteristics \textit{without ensuring} a hill-like profile. In the vortex scenario the thermomagnetic forces ge\-ne\-rate phase slippage that triggers a Nernst response with a positive sign due to the entropy carried inside the normal cores. Although there are mechanisms permitting a vortex Nernst signal even outside superconducting regions \cite{Emery}, the specific sign shortens the list of the possible materials that could belong in this category.

\begin{figure}[t]\centering
\begin{minipage}[b]{3.4in}
\includegraphics[width=0.985\textwidth]{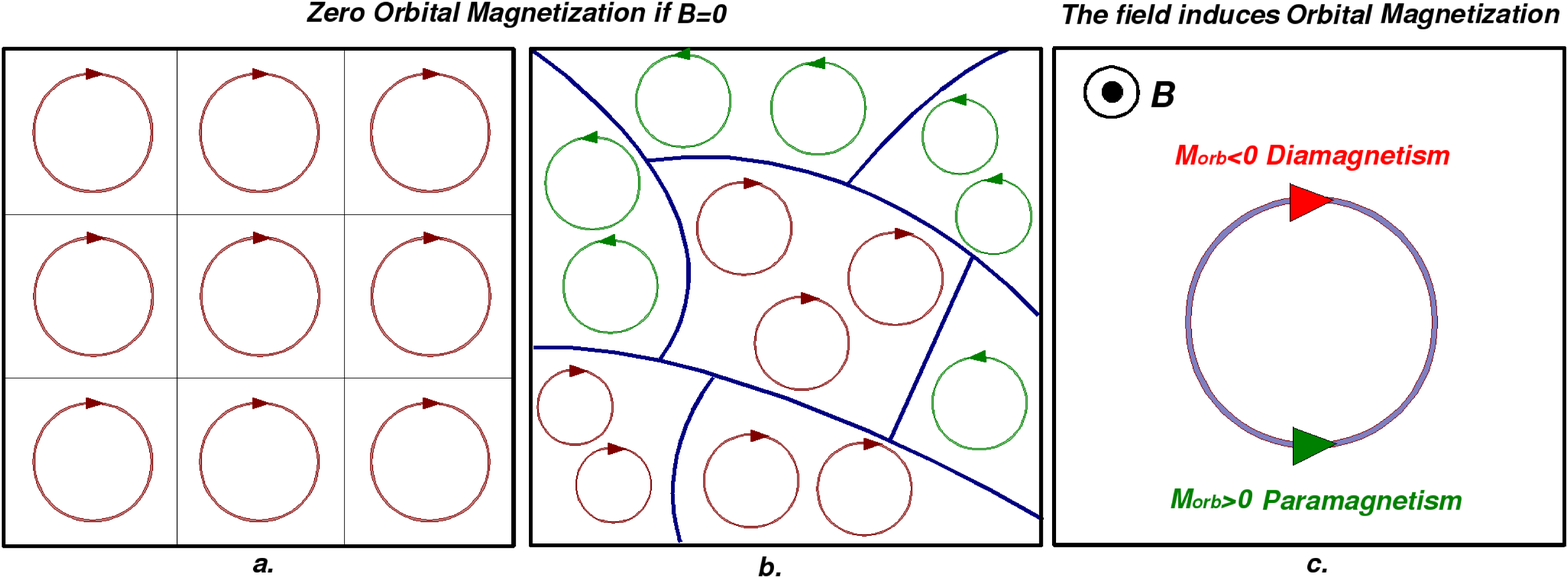}
\caption{(Color online) a. Symmetric current loops generated from a $d_{x^2-y^2}$ density wave. b. Domains of current loops in a masked chiral $d_{xy}+id_{x^2-y^2}$ density wave. c. In the previous cases time-reversal is violated only locally, while in the presence of a field, macroscopic chirality is generated accompanied by a strong paramagnetic or even diamagnetic response.}
\label{fig:1}
\end{minipage}
\end{figure}

Despite their differences, the missing link that binds together these distinct classes of materials, is the unconventional character of pairing in both superconducting and particle-hole channels. Characteristic example is the $id_{x^2-y^2}$ density wave (DDW) state, which constitutes a prominent candidate for understanding the pseudogap regime of the hole-doped cuprates \cite{Chakravarty}. This state violates time-reversal locally, since it generates a \textit{symmetric pattern of current loops} that yields a zero intrinsic orbital moment when summed all over the sample (Fig.\ref{fig:1}a). Nevertheless, the application of a magnetic field along the $z-$axis induces a $d_{xy}$ density wave component \cite{FICDDW,Balatsky}, that violates parity, destroying the symmetric distribution of the current carrying loops. The emergent $d_{xy}+id_{x^2-y^2}$ state \textit{is chiral}, since it violates time-reversal globally, generating a macroscopic current loop and a concomitant intrinsic orbital moment (Fig.\ref{fig:1}c). The field-induced chiral DDW, \textit{extends} the well established DDW picture for the pseudogap state, as it provides additional features \textit{necessary} for interpreting puzzling phenomena when an external magnetic field is present, such as the Polar Kerr effect \cite{Tewari}, the enhanced diamagnetism \cite{Meissner} and the giant tilted-hill Nernst response, discussed here.

In this Letter, we demonstrate that a chiral ground state is responsible for the enhanced Nernst response observed in a number of compounds belonging to the aforementioned classes of materials. Since, analogs of the DDW state have been also proposed for ${\rm CeCoIn_5}$ \cite{Maki}, ${\rm URu_2Si_2}$ \cite{Coleman} and the antiferromagnetic state of the iron pnictides \cite{sdwpnictides}, the emergence of chiral orders similar to the chiral DDW is anticipated, providing a unique understanding of the magnetic field anomalous response. In our picture, the giant transverse thermoelectricity is the consequence of a finite Berry curvature \cite{NiuReview}, in a manner radically different from the Anomalous Nernst effect considered earlier \cite{Zhang}. Specifically, we consider that our chiral state is in the vicinity of an insulator, which is crucial for obtaining the tilted-hill temperature profile. In our case the Hall angle is not negligible and one should keep the following full expressions for the Nernst signal ${\cal N}\equiv E_y\up/(-\partial_x\up T)=(\sigma_{xx}\alpha_{xy}-\alpha_{xx}\sigma_{xy})/(\sigma_{xx}\up\sigma_{yy}+\sigma_{xy}^2)$ and the thermopower ${\cal S}\equiv E_x\up/(\partial_x\up T)=(\alpha_{xx}\sigma_{yy}+\alpha_{xy}\sigma_{xy})/(\sigma_{xx}\up\sigma_{yy}+\sigma_{xy}^2)$.

To illustrate the chirality induced tilted-hill giant Nernst signal we shall consider the chiral DDW, earlier invoked for understanding the pseudogap regime of the underdoped cuprates \cite{FICDDW,Balatsky,Zhang,Tewari,Meissner,Partha}, while its realization with optical lattices has also been proposed \cite{OpticalLattices}. Our arguments apply to all the similar systems owing a non-zero Berry curvature. We employ the hamiltonian ${\cal H}(\bm{k})=\bm{g}(\bm{k})\cdot\bm{\tau}-\mu$, with $\mu$ a finite chemical potential and $\bm{g}(\bm{k})$ a vector in the space spanned by the isospin $\bm{\tau}$ Pauli matrices. In this case, $ g_1\up=\Delta_1\up\sin(k_x a)\sin(k_y a)$ and $g_2\up=\Delta_2\up[\cos(k_x a)-\cos(k_y a)]$ correspond to the $d_{xy}$ and $d_{x^2-y^2}$ density wave gap functions respectively, while $g_3\up=-2t[\cos(k_x a)+\cos(k_ya)]$ represents the n.n. hopping term. We must remark that the above convention for $g_{1,2}(\bm{k})$, will provide $\Delta_{1,2}>0$ when solving the self-consistency equations in the presence of an external magnetic field ${\cal B}_z$ \cite{FICDDW}. This hamiltonian has two eigenenergies $E_{\pm}(\bm{k})=-\mu\pm E(\bm{k})$ with $E(\bm{k})=\left|\bm{g}(\bm{k})\right|$. The non trivial topological content of this system generates a 
$z-$axis oriented Berry curvature \cite{FICDDW,Zhang,BerryCDDW}, defined as 
\bea\Omega_{\nu}^z(\bm{k})=-\frac{\nu}{2E^3(\bm{k})}\ph\bm{g}(\bm{k})\cdot\left(\frac{\partial\bm{g}(\bm{k})}{\partial k_x\up}\times\frac{\partial\bm{g}(\bm{k})}{\partial k_y\up}\right),\nu=\pm\,.\eea

The existence of a non-zero Berry curvature creates an effective magnetic field in the ground state leading to an anomalous Hall response. Specifically, the electric and thermoelectric Hall conductivities are both non zero even in the absence of a magnetic field and read \cite{Zhang,Tewari,BerryCDDW,NiuReview}
\bea \sigma_{xy}^{\cal B}&=&-\frac{2e^2}{v\hbar}\sum_{\bm{k},\nu=\pm}n_F\up[E_{\nu}^{\cal B}({\bm{k}})]\Omega_{\nu}^z(\bm{k})\,,\\
\alpha_{xy}^{\cal B}&=&\phd\frac{2e}{v\hbar T}\sum_{\bm{k},\nu=\pm} \left\{E_{\nu}^{\cal B}(\bm{k})n_F\up[E_{\nu}^{\cal
B}(\bm{k})]\right.\no\\&+&\left.k_B\up T\ln\left(1+e^{-E_{\nu}^{\cal B}(\bm{k})/k_B\up T}\right)\right\}\Omega_{\nu}^z(\bm{k})\,,\eea

\noi with $n_F\up$ the Fermi-Dirac distribution and $v$ the number of the lattice sites. The longitudinal electric and thermoelectric conductivities are calculated within the Boltzmann approximation, based on the equations 
\bea \sigma_{xx}^{\cal B}
&=&-
\frac{2e^2}{v\hbar^2}\sum_{\bm{k},\nu=\pm}n_F'\left[E_{\nu}^{\cal B}({\bm{k}})\right]\frac{v_{x,\nu}^2(\bm{k})\tau_{\nu}\up(\bm{k})}{\left[1+\frac{e}{\hbar}{\cal B}\Omega_{\nu}^z(\bm{k})\right]^2}\,,\\
\alpha_{xx}^{\cal B}
&=&\frac{2e}{v\hbar^2T}\ph\sum_{\bm{k},\nu=\pm}n_F'\left[E_{\nu}^{\cal
B}({\bm{k}})\right]\frac{v_{x,\nu}^2(\bm{k})E_{\nu}^{\cal
B}(\bm{k})\tau_{\nu}\up(\bm{k})}{\left[1+\frac{e}{\hbar}{\cal B}\Omega_{\nu}^z(\bm{k})\right]^2}\,,\quad\eea

\noi with the prime denoting differentation. We have introduced the field dependent dispersions
$E_{\nu}^{\cal B}(\bm{k})=E_{\nu}(\bm{k})-\bm{m}(\bm{k})\cdot\bm{{\cal B}}$ and the zero field velocities
$\bm{v}_{\nu}\up(\bm{k})=\bm{\nabla_k}\up E_{\nu}\up(\bm{k})$. We have also incorporated the Berry phase correction to the longitudinal conductivities through the modified DOS and relaxation time $\tau_{\nu}\up(\bm{k})\equiv\tau$ \cite{NiuReview}. In addition, we have considered only the orbital coupling with the field originating from the intrinsic orbital moment defined as $\bm{m}(\bm{k})=\frac{e}{\hbar}E(\bm{k})\bm{\Omega}_+\up(\bm{k})$. For the energy scales presented here the Zeeman interaction is negligible while for stronger fields we expect it to become significant enough to destroy the chiral state. For the numerical results we have employed parameters relevant for the underdoped cuprates: $t=250 {\rm meV}$, $\mu=0$, $a=5${\AA}, $\Delta_2\up=53{\rm meV}$ and $\Delta_1\up({\cal B})=1.76+4.1\sqrt{{\cal B}}\ph{\rm meV}$, a dependence previously extracted self-consistently in Ref.\cite{FICDDW}. The relaxation time was set $\tau=10^{-13}s$, i.e. of the same order inferred from resistivity measurements \cite{Green}. Notice that in stark contrast to the quasi-particle Nernst signal, chirality driven Nernst response \textit{does not depend linearly on the re\-la\-xa\-tion time}, since $\sigma_{xy}^{\cal B}$ and $\alpha_{xy}^{\cal B}$ have a topological origin.

\begin{figure}[t]\centering
\begin{minipage}[b]{3.4in}
\includegraphics[width=0.475\textwidth]{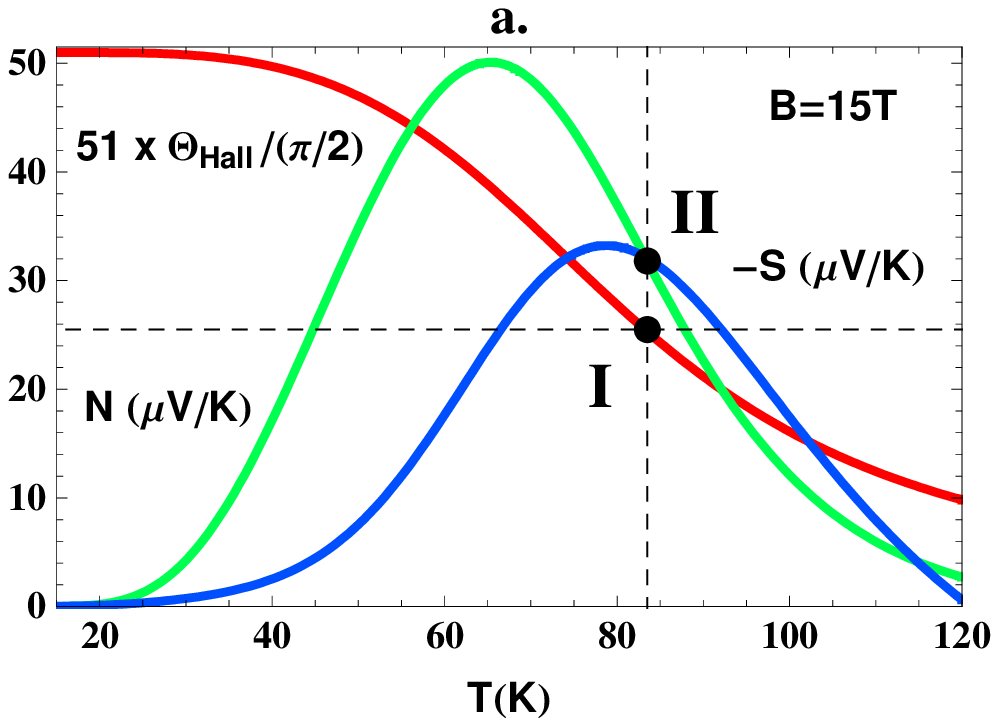}
\hspace{0.05in}
\includegraphics[width=0.475\textwidth]{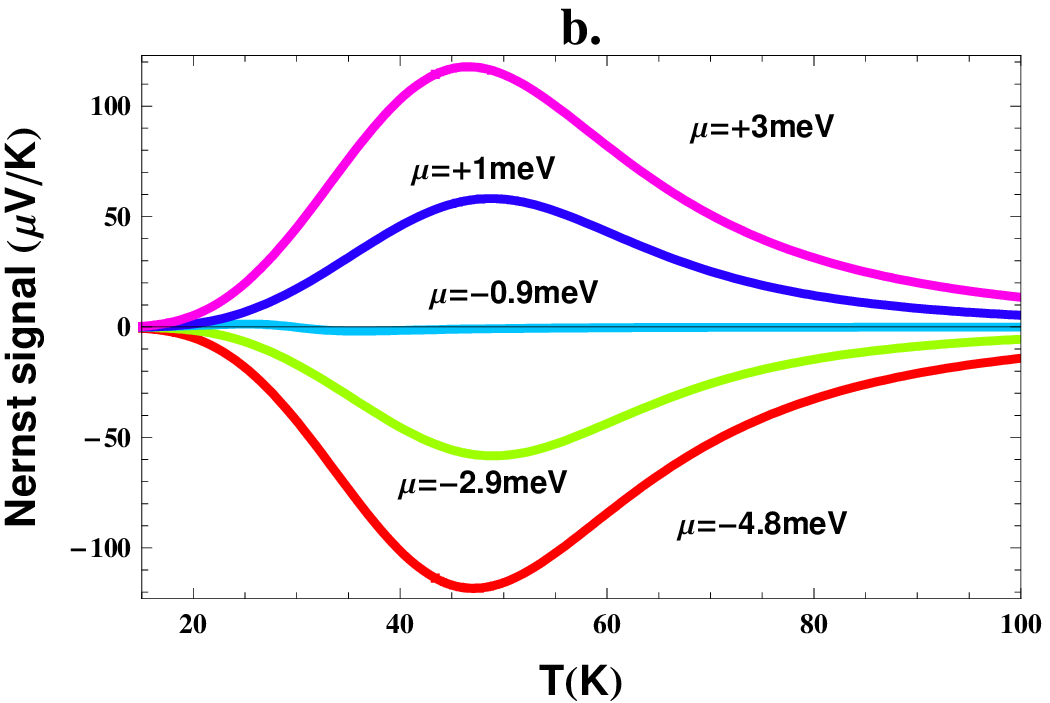}
\caption{(Color online) a. For low temperatures the Nernst signal and thermopower increase exponentially while the Hall angle is large. At the crossing point (${\rm I}$) where $\sigma_{xx}=\sigma_{xy}$, the Nernst signal equals the thermopower (${\rm II}$) while their sign may be the same or opposite. At this crossover regime, the usually large magnitude of the thermopower ensures that the Nernst signal will be enhanced since they necessarily become equal. Just after the crossing temperature, $\sigma_{xx}>>\sigma_{xy}$ forces the Nernst signal to decrease rapidly, giving rise to the tilted-hill Nernst profile. b. Nernst signal dependence on doping. Tuning of the chemical potential inverts the Nernst signal's sign. The Nernst signal is symmetric around an offset $\mu\simeq-0.8{\rm meV}$ needed to compensate the term $-m_z{\cal B}$ (${\rm {\cal B}=5T}$).}
\label{fig:2}
\end{minipage}
\end{figure}

As expected from the strongly insulating character of our state, we obtain a \textit{large constant Hall conductivity for all the temperatures considered here}. In the same time, the longitudinal conductivities grow very slowly with an exponential tempe\-ra\-tu\-re law, while the Hall angle $\tan\Theta_H\up\equiv\sigma_{xy}/\sigma_{xx}$, is very large. In this temperature region, the thermopower and Nernst are provided by the approximate relations ${\cal S}\simeq\alpha_{xy}/\sigma_{xy}$ and ${\cal N}\simeq-\alpha_{xx}/\sigma_{xy}$. Consequently they both increase exponentially for low temperatures, as the only contributions to $\alpha_{xy}$ and $\alpha_{xx}$ come from thermally excited quasiparticles, producing \textit{the left half of the hill profile} (Fig. 2a). Upon raising the temperature, we pass from the large Hall regime to the one dominated by significant dissipative longitudinal transport. In this case, the usual expressions ${\cal S}=\alpha_{xx}/\sigma_{xx}$ and ${\cal N}=\alpha_{xy}/\sigma_{xx}$ apply. In the course from high to low Hall angle, the system unavoidably crosses the $\sigma_{xy}=\sigma_{xx}$ point for a specific temperature (Point ${\rm I}$ in Fig.\ref{fig:2}a). At this crossing point there are three possible scena\-rios, that depend on the ratio of the thermoelectric conductivities. When $\alpha_{xy}\up>>\alpha_{xx}\up$ ($\alpha_{xy}\up<<\alpha_{xx}\up$) the thermopower and the Nernst signal have the same (opposite) sign while their magnitudes become equal to $|\alpha_{xy}/2\sigma_{xx}|$ ($|\alpha_{xx}/2\sigma_{xx}|$) (Point ${\rm II}$ in Fig.\ref{fig:2}a). Taking into account that the chiral state is enhanced by the field and the Hall conductivity remains approximately constant, we understand that the behaviour of the Nernst signal and thermopower are mainly determined by the thermoelectric coefficients. Finally, if $\alpha_{xy}\simeq\alpha_{xx}$, the Nernst signal va\-ni\-shes.

\begin{figure}[t]\centering
\begin{minipage}[b]{3.4in}
\includegraphics[width=1\textwidth]{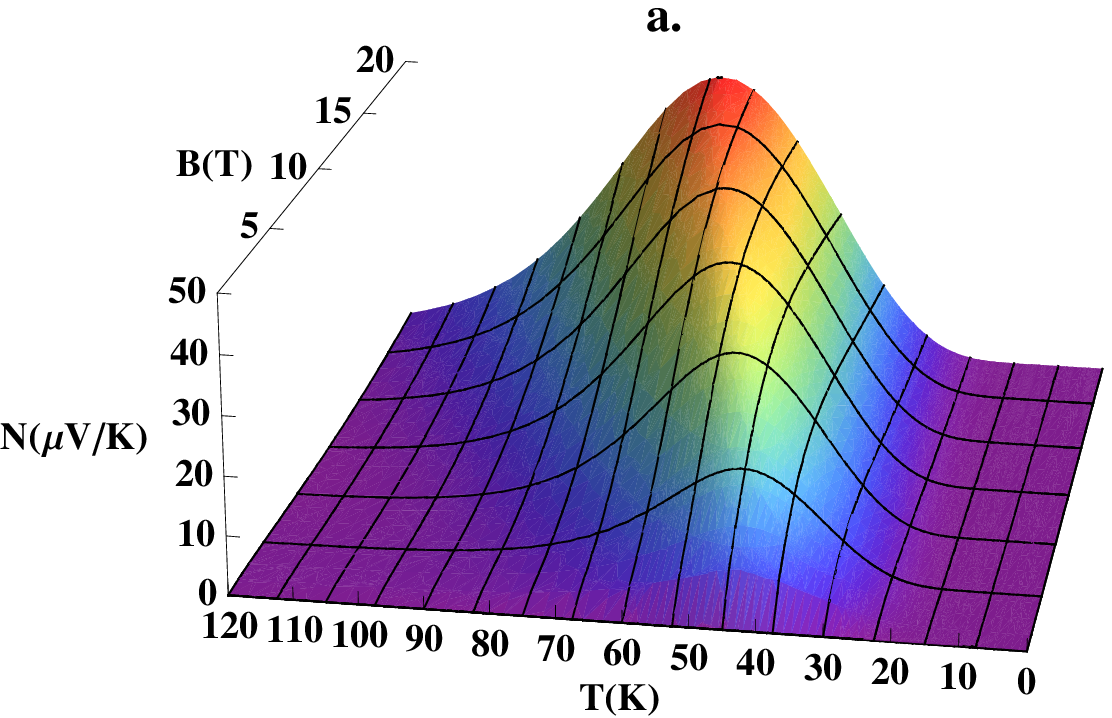}\\\vspace{0.45in}
\includegraphics[width=0.48\textwidth]{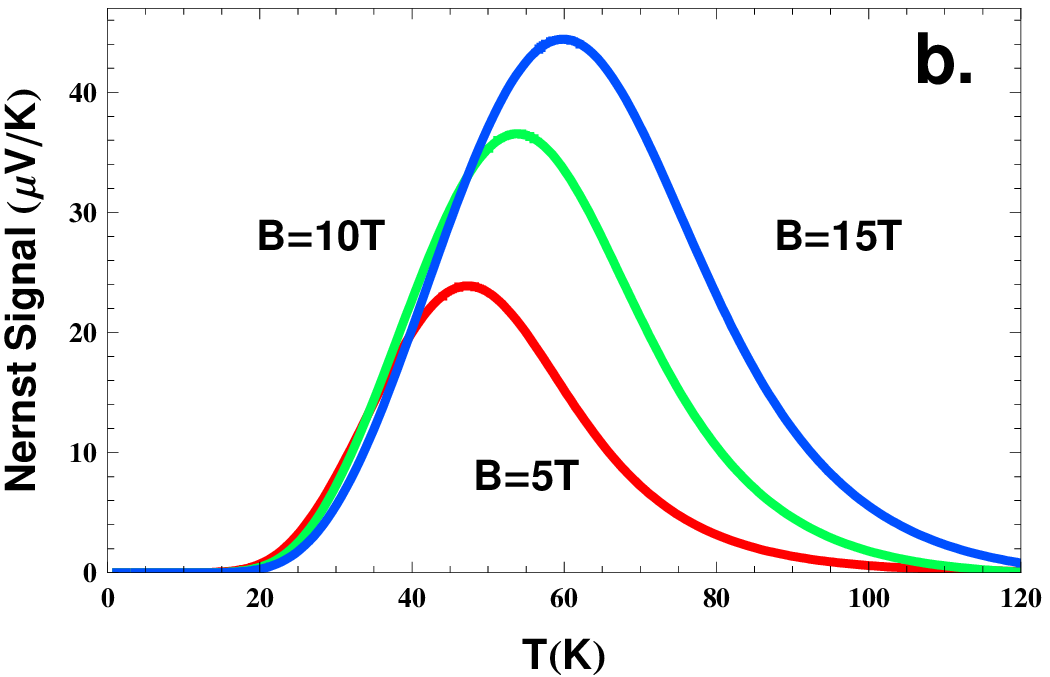}
\hspace{0.05in}
\includegraphics[width=0.48\textwidth]{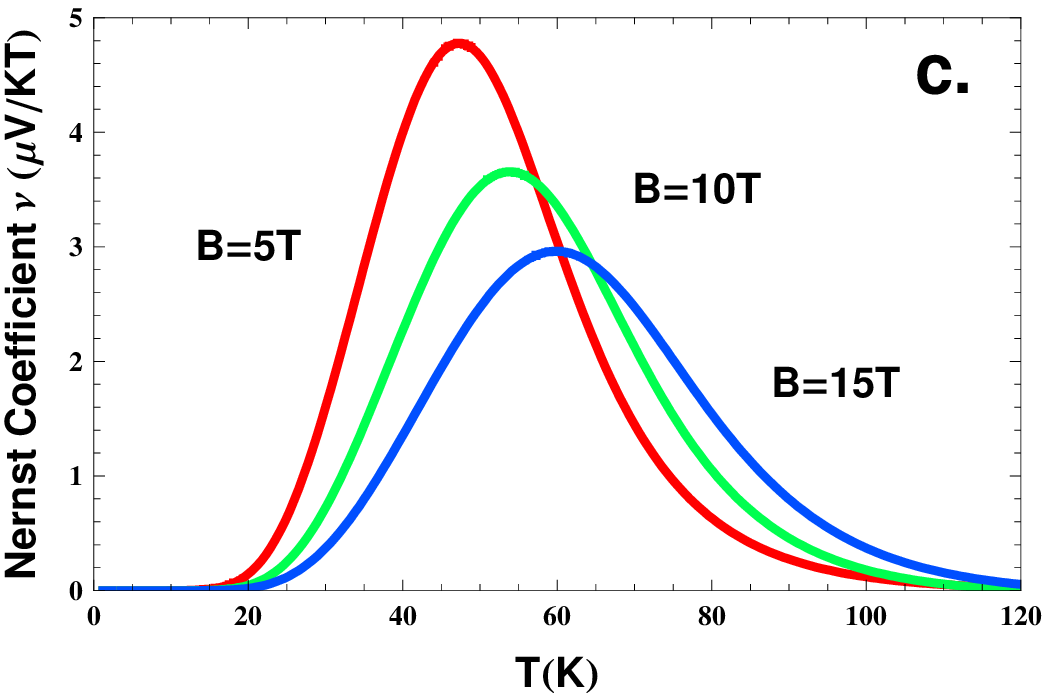}
\end{minipage}
\caption{(Color online) a. Magnetic field evolution of the tilted-hill Nernst signal temperature profile. The chirality driven Nernst response may reach the enormous values of 50$\mu V/K$ in the high field limit. b. The magnetic field enhances the Nernst signal while it shifts the Nernst peak to high temperatures. c. The Nernst coefficient decreases upon raising the field.}
\label{fig:3}
\end{figure}

In the first two cases, the emergence of a large number of excited quasiparticles strongly enhances the initially exponentially increasing $\alpha_{xx}$ and $\alpha_{xy}$, opening the channel of the giant Nernst signal. \textit{As a matter of fact, the key that unlocks the anomalous thermoelectricity is the existence of this crossing point at which the Nernst signal becomes equal to the thermopower}. In this manner the usually large magnitude of the thermopower, carries the Nernst response along to high values. However, very soon after the giant Nernst signal has been built up, the dominance of the Hall conductivity is ruined. This suppresses again the Nernst signal, leading to the characte\-ri\-stic tilted-hill temperature profile. In addition to this intricate temperature evolution, there is also a delicate interplay of the signal's sign upon doping. In Fig.\ref{fig:2}b we demonstrate that \textit{tuning a chemical potential can controllably invert the sign of the Nernst response}. The sign inversion is a straightforward consequence of the linear dependence of the thermoelectric conductivities $\alpha_{xx}$ and $\alpha_{xy}$ on doping. The arising magnetic field evolution of the tilted-hill temperature profile is presented in Fig.\ref{fig:3}, with a magnitude that reaches the enormous values of $50\mu V/K$ in the high field limit. As it is apparent from Fig.\ref{fig:3}b, \textit{the magnetic field enhances the signal while it also shifts the temperature of the Nernst peak to higher values}. In contrast, as we notice in Fig.\ref{fig:3}c, the Nernst coefficient \textit{diminishes} with the increase of the magnetic field something that is in full accordance with the experimental findings in the cuprates. 



\begin{figure}[t]\centering
\includegraphics[width=0.42\textwidth]{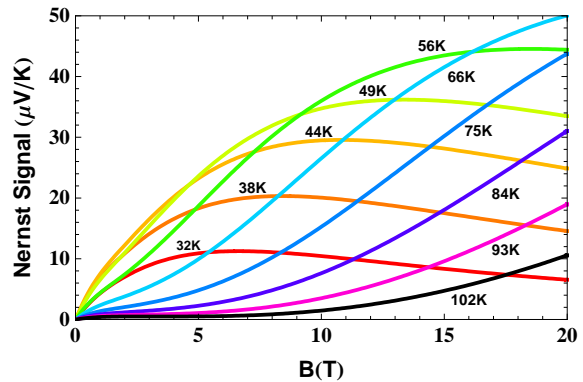}
\caption{(Color online) Magnetic field dependence of the Nernst response for different temperatures. The Nernst signal demonstrates a strong non-linearity for T$<$56K that is qualitatively very similar to the data of underdoped ${\rm La_{1.93}Sr_{0.07}CuO_4}$ for T$<$12K. The slope change after a critical temperature (T$=$56K here), is consistent with the experimental findings.}
\label{fig:4}
\end{figure}

As far as the magnetic field dependence of the Nernst signal is concerned, Fig.\ref{fig:4} shares some common features with the available experimental data for underdoped ${\rm La_{2-x}Sr_{x}CuO_4}$ (LSCO) and ${\rm Bi_2Sr_2CaCu_2O_{8+\delta}}$ (Bi2212) with hole densities $x=0.07$ and $x\sim 0.09$ respectively \cite{cuprates}. The temperature scales of the Nernst signal presented here, fit nice with the scales encountered in Bi2212. Note that the Nernst signal in our case is more than one order of magnitude larger than the one observed. The field evolution presented here for 40K$<$T$<$50K matches well the experimental in the same interval for Bi2212. Moreover, there is a qualitative fit between the range T$<$12K for LSCO and our case for T$<$56K. In this range, the Nernst signal is strongly non-linear. For T$>$56K our Nernst signal starts decreasing accompanied by a change in the curve's slope. This is a behavior generally compatible with the experimental observations. However, the emerging field dependence does not fit very well with the linear trend observed in experiments. This discrepancy could originate from not having taken into account the quasiparticle contributions to $\sigma_{xy}^{\cal B}$ and $\alpha_{xy}^{\cal B}$. These contributions depend linearly on the magnetic field and we expect them to have a ``linearizing'' impact on the Nernst signal evolution, especially in the high-field case.  

Our picture is also fully compatible with the observed enhanced diamagnetism \cite{cuprates} in the non-superconducting pseudogap regime of the hole-doped cuprates. In fact, the chiral DDW may exhibit strong diamagnetism that in the ideal case becomes a Meissner effect \cite{Meissner}. In the specific case reported here, we have verified that the giant positive Nernst signal is accompanied by an enhanced orbital diamagnetism. This implies that the presence of \textit{chiral states could have been misre\-co\-gni\-sed as vortices} in se\-ve\-ral materials, as in many cases they combine enhanced diamagnetism and Nernst response. Finally, our arguments find further support in the recent tantalizing observations \cite{StripesEnhanceNernst,Chang,NematicNernstCuprates}, demonstrating that at least for some cuprate superconductors, employing a vortex Nernst scenario is not essential for a large Nernst signal, as this can be enhanced by striped or nematic phases. A chiral DDW is certainly in accordance with these findings, in terms of a \textit{striped} or a \textit{nematic} chiral DDW, with the latter state arising from the coexistence of a chiral DDW with a Pome\-ra\-nchuk state \cite{Pomeranchuk}. 

In conclusion, we have proposed a novel mechanism for generating a tilted-hill giant Nernst signal that could shed light on the connection of anomalous thermoelectricity in a plethora of materials, including high-Tc superconductors. From our point of view, the Nernst signal emerges from a magnetic-field-induced chiral order due to the presence of an unconventional particle-hole condensate. The arising Nersnt signal has a non linear temperature and magnetic field dependence while it is enhanced upon increasing the field. The induced chiral states responsible for the anomalous thermoelectricity may also demonstrate enhanced diamagnetism, properties so far considered as definite signatures of vortex Nernst signal. Nevertheless, the origin of the Nernst signal may be distinguished in this case. The vortex Nernst signal is always positive while a chirality induced thermoelectric response changes sign by varying doping. Our results indicate that the giant Nernst effect in various materials not only shares a common chiral origin but it may be also considered as the fingerprint for the existence of unconventional ordered states such as the d-density wave in the cuprates.

P.K. is indebted to K. Behnia for valuable information on the Nernst response in heavy fermions. Moreover, we are grateful to C. Panagopoulos, P. Thalmeier, A. Aperis and S. Kourtis for comments and stimulating discussions. P.K. also acknowledges financial support by the Greek National Technical University Scholarships Foundation.

\end{document}